# β-Ag$_2$Te: A topological insulator with strong anisotropy

Azat Sulaev[1+], Peng Ren[1+], Bin Xia[1], Qing Hua Lin[1], Ting Yu[1], Caiyu Qiu[1], Shuang-Yuan Zhang[2], Ming-Yong Han[2], , Zhi Peng Li[3], Wei Guang Zhu[3], Qingyu Wu[4], Yuan Ping Feng[4], Lei Shen[4]*, Shun-Qing Shen[5], and Lan Wang[1] *

1. School of Physical and Mathematical Science, Nanyang Technological University, Singapore, 637371, Singapore
2. Institute of Materials Research and Engineering, Singapore, 117602, Singapore
3. School of Electronics and Electrical Engineering, Nanyang Technological University, Singapore, 639789, Singapore
4. Department of Physics, National University of Singapore, Singapore
5. Department of Physics, The University of Hong Kong, Pokfulam Road, Hong Kong, People's Republic of China

We present evidence of topological surface states in β-$Ag_2Te$ through first-principles calculations and periodic quantum interference effect in single crystalline nanoribbon. Our first-principles calculations show that β-$Ag_2Te$ is a topological insulator with a gapless Dirac cone with strong anisotropy. To experimentally probe the topological surface state, we synthesized high quality β-$Ag_2Te$ nanoribbons and performed electron transport measurements. The coexistence of pronounced Aharonov-Bohm oscillations and weak Altshuler-Aronov-Spivak oscillations clearly demonstrates coherent electron transport around the perimeter of β-$Ag_2Te$ nanoribbon and therefore the existence of metallic surface states, which is further supported by the temperature dependence of resistivity for β-$Ag_2Te$ nanoribbons with different cross section areas. Highly anisotropic topological surface state of β-$Ag_2Te$ suggests that the material is a promising material for fundamental study and future spintronic devices.

+These authors contributed equally to this work

*Email: shenlei@nus.edu.sg, wanglan@ntu.edu.sg

Topological insulator is a state of quantum matter characterized by $Z_2$ invariance. It is composed of a fully filled insulating bulk state and an odd number of massless spin-helical Dirac cones of two-dimensional surface states[1-3]. Due to the fascinating new physics and great potential application in spintronics and quantum computation, topological insulator quickly becomes a trend research field in condensed matter physics. Various two- and three-dimensional topological insulator systems have been proposed via band structure calculation. However, up to date, only CdTe/HgTe/CdTe[4,5] and InAs/GaSb[6] quantum well structure have been experimentally confirmed to be two-dimensional topological insulator while strained HgTe[7,8] and several Bi based compounds, such as $Bi_xSb_{1-x}$, $Bi_2Se_3$, $Bi_2Te_3$, *etc* [9-20], have been confirmed to be three-dimensional topological insulators. Thus, exploiting and identifying other good candidates of topological insulators with special characteristics is crucial and desirable.

β-$Ag_2Te$, a narrow band gap nonmagnetic semiconductor, shows unusual large and nonsaturating quasi-linear magnetoresistance in the field range of $10$-$10^5$ Oe and the temperature range of 5-300 K[21,22]. The origin of the unusual magnetoresistance has generated debates since its discovery[23-25]. Recently, β-$Ag_2Te$ was predicted to be a topological insulator with gapless Dirac-type surface states via band structure calculation[26]. It was proposed that the observed unusual magnetoresistance may largely come from the surface or interface contribution. The characteristic feature of this new binary topological insulator is a highly anisotropic Dirac cone, which is very different from other known topological insulators as aforementioned. Physical realization of topological insulator with highly anisotropic Dirac fermions may lead to a discovery of novel electronic states and long spin relaxation time in topological insulators[27]. The long spin relaxation time is extremely important for the application of topological insulator in spintronics, which has barely been studied so far.

In order to verify the essential features of the aforementioned picture, we utilized a different method to perform band structure calculations and carried out magnetotransport experiments on single crystalline β-$Ag_2Te$ nanoribbons. Our calculations confirmed the anisotropic topological surface state, although the shape of the surface state of our calculations is different from that in the literature[26]. The existence of the surface states in $Ag_2Te$ is confirmed experimentally for the first time, based on the Aharonov-Bohm interference pattern obtained in the magnetotransport measurements at low temperatures. It is further supported by the temperature dependence of the resistivity of nanostructure with different sizes. Due to a phase transition at 145 °C for $Ag_2Te$ compound, it is very difficult to synthesize high quality bulk single crystalline β-$Ag_2Te$. Up to date, all the previous transport measurements were performed using polycrystalline bulk samples or thin films. To the best of our knowledge, our transport measurement in fact is the first electrical transport measurement on single crystalline β-$Ag_2Te$.

**Results**

**First-principles calculations.** In the band structure calculations of Zhang *et al*[26], first-principles calculations were utilized in the bulk state of β-$Ag_2Te$, while tight-binding method was employed to calculate the surface state. To keep the calculations consistently and improve the accuracy of description of the surface state, we used first principles calculations for both bulk and surface band-structure calculations.

Fig. 1 shows the calculated surface band structure of *β*-Ag$_2$Te with spin-orbital coupling (SOC) included and the out-of-plane of the surface is along the *c*-axis. The shaded area is the bulk band structure projected to the 2D Brillouin zone. It is seen that the bulk gap is around 0.1 eV which is in good agreement with experimental data[28, 29]. The upper inset shows the shape of the 2D Brillouin zone for the surface unit cell. The red solid lines are the surface topological state, which show a single Dirac cone on the surface, which is similar to that obtained via tight banding method[26] although the shapes of the surface state are different.

The bottom inset shows the constant-energy contours which cut through the surface state below the Dirac point at energies of -10 meV, -20 meV and -30 eV. The red (blue) color means the out-of-plane components pointing out-(in-) ward of the plane. The color scale, with blue and red, indicates the intensity of negative and positive values representing the projection $S_z$. Our results show the spin direction of the surface states processes counterclockwise around the Γ point below the Dirac point and has out-of-plane component. The most interesting phenomenon is the anisotropic surface Dirac cone of β-Ag$_2$Te which is different from the round and warped hexagonal surface state of Bi$_2$Se$_3$ and Bi$_2$Te$_3$[10, 30]. It is because of the absence of the rotational symmetry in β-Ag$_2$Te.

**Sample structure.** Ag$_2$Te nanoribbon**s** were synthesized via vapor deposition methods on sapphire substrates. As grown ribbons are typically 100-200 nm thick, 100 nm to several micrometers wide, and up to tens of micrometers long. The transmission electron microscopy (TEM) and corresponding selected area electron diffraction (SAED) investigation was carried out for the as-prepared nanoribbons, which were separated from the sapphire substrate by ultrasonication. Figure 2a shows a TEM image of nanoribbons. The diffraction pattern (Fig. 2a inset) shows clearly that the nanoribbon is single crystalline and matches well with the standard monoclinic β-Ag$_2$Te (ICSD file number: 073402) single crystal electron diffraction pattern (Fig. 1b) at zone axis [4 $\bar{8}$ 5]. The electron diffraction pattern are simulated based on monoclinic Ag$_2$Te single crystal with cell parameters $a$ = 8.164 Å, $b$ = 4.468 Å, $c$ = 8.977 Å, space group P2$_1$/c. A high resolution TEM picture of focused ion beam milled β-Ag$_2$Te is shown in Fig. 2c. It is clear from the picture and corresponding fast Fourier transform pattern that our single crystallineAg$_2$Te is very high quality and the growth direction is perpendicular to the (1$\bar{2}$1) plane direction. Energy-dispersive X-ray spectroscopy shows that the atomic ratio of Ag and Te is 2.02 (with an error bar of 0.5%) as shown in Fig. S1 and S2 (supplementary information). The distribution of Ag and Te is homogeneously in the sample, as shown by the EDS mapping in Fig. 2d.

**Aharonov-Bohm interference.** The magnetoresistance measurement results for three samples (A, B and C) are presented in this paper. The cross section areas of sample A, B and C are 6.31 width (W) × 0.211 thickness (T) μm$^2$, 1.11(W) × 0.102 (T) μm$^2$ and 0.191(W) × 0.098(T) μm$^2$, respectively. Fig. 3a shows the schematic diagram of the standard four point measurements used in our experiments. Fig. 3b shows a scanning electron microscopic image of a typical device. A device composed of a nanoribbon and five equal-distance separated gold contacts are shown in the image. To probe the quantum interference effect in the β-Ag$_2$Te nanoribbons, we measured the magnetoresistance of sample C with the applied magnetic field parallel to the current flowing direction as shown in the schematic diagram (Fig. 4d). As shown in Fig. 4a and 4b, pronounced and reproducible resistance oscillations with a period of 0.227 Tesla are observed at 2 K, 4 K, 6 K, 8 K and 10 K. The oscillation amplitude at 2 K is about 1.5% of the total resistance. As indicated by the arrows in Fig. 4a and 4b, there are also weak oscillations with a period of 0.113 Tesla overlapped on the pronounced oscillations with the period of 0.227 Tesla. At 2 K, the measurements were performed with a magnetic field ramping from 0 to 9 Tesla and then ramping back from 9 to 0 Tesla. At other temperatures, the measurements were carried out up to 2.5 Tesla. It is found with no surprise that the sweeping direction has no effect on the resistance oscillations. This kind of periodic magnetoresistance oscillation can only be induced by quantum interference Aharonov-Bohm (A-B) effect or Altshuler-Aronov-Spivak (AAS) effect. As the cross section of the nanoribbon under measurement is $S$ = thickness (98 nm) × width (191 nm), the corresponding A-B and AAS oscillation periods should be 0.221 Tesla and 0.110 Tesla, respectively. Considering the experimental error bar, we can therefore conclude that the oscillations in our measurements are attributed to a strong A-B effect and a weak AAS effect. Prior to the work herein, the A-B oscillation was observed in $Bi_2Se_3$ and $Bi_2Te_3$, which was regarded as one of the straightforward evidences for topological insulators[31-33]. Theories of quantum oscillations for topological insulator include two types, the ballistic transport and diffusive transport[34, 35]. As a $2\pi$ rotation of a spin around the curved surface of a topological nanoribbon generates a Berry phase of $\pi$, the magnetoresistance of an undoped ballistic nanoribbon is expected to oscillate with a period of one magnetic flux quantum h/e and a resistance minimum at half flux quantum $\phi$ = h/2e (a maximum at zero flux). For topological insulator nanoribbons with strong disorder, electron transport is diffusive. Quantum correction is caused by the interference between time reversal paths, and the resistance oscillates with a period of h/2e. The magnitude of spin-orbit interaction determines appearance of the maximal or minimal resistance at zero flux. For $Ag_2Te$, the A-B oscillation shows a minimal resistance at zero flux and strong h/e and weak h/2e periodicity, which cannot be explained by the aforementioned two scenarios. Recent theoretical simulations[34, 35] proposed that the disorder and the Fermi level position can determine the oscillation period and whether the resistance has a minimum or maximum at zero flux. With weak antilocalization and strong disorder induced diffusive motion, the resistance shows h/2e period and has a minimum at ϕ = h/2e. For samples with weak disorder, the resistance of the h/e period has a minimal value at ϕ = 0 or ϕ = h/2e depending on the Fermi level. Based on theoretical simulation results, we speculate that our $Ag_2Te$ nanoribbon has a mediate disorder, and therefore exhibit oscillations with both h/e and h/2e periods. The other possibility is that the surface state includes a dominating topological nontrivial part and a weak topological trivial part coming from band bending at the surface, which shows A-B oscillations with a period of h/e and AAS oscillation with a period of h/2e, respectively.

The observation of A-B oscillations with a period of h/e unambiguously suggests the existence of conducting surface states in $Ag_2Te$ nanoribbon, and the existence of conducting surface states on all the crystalline orientations (top, bottom and sides) provides explicit evidences for the topological origin of the surface states as expected by a topological insulator. The topological trivial surface states usually show an AAS oscillation strongly depend on the surface orientation due to bonding orientation. The dominance of the A-B effect (period of h/e), instead of the normal AAS effect (period of h/2e), is a fingerprint of topological surface states with weak disorder, which is related to the Dirac type Hamiltonian of the helical surface state. Recently, quantum oscillations with a period of h/e were observed in InN nanowires, which have a topological trivial surface state due to band bending[36]. However, the oscillation is caused by energy level splitting, which can only be observed in ballistic electron transport. The oscillation magnitude will also not decay with increasing magnetic field. The electron transport in our experiment is definitely diffusive as the interval between the electrodes is ~1.5 μm. The magnitude of oscillations also decays with increasing magnetic field, which is a standard characteristic of quantum interference induced oscillation. Therefore, the A-B oscillation (h/e period) in our measurements should be a finger print of topological surface state and not be the type of oscillations in InN nanowire. The decaying speed of A-B or AAS oscillation with increasing magnetic field is related to the thickness of the surface states. The decay of the oscillation magnitude with increasing magnetic field in $Ag_2Te$ is faster than that in $Bi_2Se_3$, which may indicate thicker surface states in $Ag_2Te_3$. In Fig. 4c, the temperature dependence of the amplitude of the A-B oscillation is plotted and fitted using a T power law. The oscillation amplitude roughly scales with $T^{-0.46}$ between 2 K and 10 K. Similar behaviors were also found in the A-B effect of $Bi_2Se_3$ and $Bi_2Te_3$ nanowire, which is attributed to the temperature dependence of the phase coherence time, $\tau_\phi \sim h/k_BT$. The oscillation amplitude is proportional to the phase coherence length $L_\phi = (D\tau_\phi)^{-1/2} \sim T^{-1/2}$, which agrees with the experimental results.

**Temperature dependence of resistance.** To further study the effect of topological surface state in electron transport, we also measured the temperature dependence of resistivity of $\beta$-$Ag_2Te$ single crystalline nanoribbons with various cross sections to find the relationship between the resistivity and cross section area. For topological insulators with the same chemical composition, due to the surface states, the resistivity should decrease with decreasing cross section at low temperatures, which has been reported in our measurements on $Bi_{1.5}Sb_{0.5}Se_{1.8}Te_{1.2}$ bulk single crystals and nanoflake devices[37]. However, we cannot figure out a clear relationship of this kind in our measurements for $\beta$-$Ag_2Te$ nanoribbons. We believe that this is caused by the small composition variation for different samples, which is a common problem for the CVD grown nanostructures, although the atomic ratio of Ag/Te determined by energy dispersive X-ray spectroscopy (EDS) is always very near 2 for all the nanoribbons measured. As $\beta$-$Ag_2Te$ is a narrow band semiconductor, slight composition change may generate different resistivity. The resistivity of our samples at 10 K with zero field varies from several mΩ·cm to thirty mΩ·cm. Although there is no clear relationship between the resistivity and the area of the cross-section of samples, the shape of the resistivity *vs*. temperature curves does connect with the cross-section area of nanoribbons. The temperature dependence of normalized resistance of three β-$Ag_2$Te nanoribbons (A, B and C) with different cross-section area is shown in Fig. 5a, 5b and 5c. The three samples are prepared in one batch of growth. Sample A (the sample with the largest cross section area), shows semiconductor behavior in the measured temperature from 300 K to 10 K while both sample B and sample C show semiconductor characteristics at a high temperature region and present a semiconductor-to-metal transition at 50 K and 75 K, respectively. Sample C is also used for the A-B effect measurements (Fig. 4), which has the smallest cross section area. The decrease of the resistance with decreasing temperature is more pronounced in sample C comparing with that in sample B. As the A-B oscillation suggests the existence of topological surface states on β-$Ag_2$Te, it is very natural to use the conducting surface states to account for the size dependence of electrical transport behaviors. Since the transport contribution from the metallic surface increases with decreasing size of the sample, the metallic transport characteristic starts at higher temperatures and is more pronounced at a low temperature region.

**Discussion**

Our first-principles calculations confirm the anisotropic topological surface state in β-$Ag_2$Te. Pronounced A-B oscillations and weak AAS oscillations are observed in the magnetoresistance measurements with the magnetic field applied along the current flowing direction in a single crystalline β-$Ag_2$T nanoribbon, which provides an experimental evidence for the existence of the topological surface state. The temperature dependence of resistance provides further evidence of the topological surface states in β-Ag$_2$Te. The anisotropic Dirac cone of Ag$_2$Te suggest that it may be a promising material for fundamental research and for future spintronic devices. Based on the first electron transport experiments of single crystalline β-Ag$_2$Te, we can design many further experiments to explore the special characteristics of the topological surface states of Ag$_2$Te. For example, based on our experimental results, we believe that single crystalline thin film β-Ag$_2$Te can be grown on sapphire substrate. Using the thin film β-Ag$_2$Te, we can perform angle resolved photo emission spectroscopy (ARPES)[9-12] and in-plane anisotropic magnetoresistance measurements, which can definitely reveal the characteristics of the special highly anisotropic Dirac cone. Based on the long spin diffusion length induced by the highly anisotropic Dirac cone and colossal magnetoresistance, various novel spintronic devices can be designed and fabricated.

**Methods**

**Density Functional calculations:** Our density functional calculations were carried out using the Vienna *ab initio* simulation package (VASP)[38] with projector-augmented-wave potentials[39] and the Perdew-Burke-Ernzerhof generalized gradient approximation[40] for exchange-correlation functional. The Hamiltonian contained the scalar relativistic corrections and the spin-orbit coupling was taken into account by the second variationmethod[41]. The lattice constants of *β*-Ag$_2$Te were adopted from experiments[42]. The generic *β*-Ag$_2$Te substrate was modeled by a slab of 48 atomic layers or 12 unit cell layers along the hexagonal *c*-axis[26]. The slab calculations presented in this work were performed using symmetric setups, so the upper and lower surfaces were identical and their energy bands were degenerated. The vacuum layers were over 20 Å thick to ensure decoupling between neighboring slabs. The cutoff for plane-wave expansion was set to be 300 eV and a 7 x7 x 1 k-point mesh was used for the surface unit cell.

**Ag$_2$Te nanoribbon synthesis.** Ag$_2$Te nanoribbons were synthesized by CVD method[43] inside a 30 cm horizontal tube furnace equipped with 3 cm diameter quartz tube. The base pressure of the system was $3 \times 10^{-7}$ torr. The powder of Ag$_2$Te (Sigma-Aldrich, 99.99%) was used as a precursor placed at the hot center region. Sapphire substrates were placed downstream 14-17 cm away from the hot center region without any no catalyst. Before the usage, the substrates were cleaned using acid, ethanol and IPA. The tube was flushed three times with the argon gas to decrease oxygen contamination. The typical growth conditions for the nanoribbons are: 0.026g Ag$_2$Te, pressure 7.5 torr, source temperature 980°C, growth time 1 hour, and gas flow rate 50 s.c.c.m. Nanoribbons generally grow at the position range of 15-16 cm away from the center.

**Material characterization.** Transmission electron microscopy (TEM) images were examined by Philips CM300 field emission gun transmission electron microscope operating at an accelerated voltage of 300 kV. The deposited nanoribbons on the sapphire substrate were placed in a vial containing 2 mL methanol solvent, and subsequently ultrasonicated for 10 min to separate the nanoribbons from substrate and disperse them into solvent. A drop of such dispersion (2μL) was withdrawn and added onto a carbon-coated copper grid, which was placed on a piece of tissue paper. The solvent on the copper grid was immediately adsorbed by the tissue paper and ready for the TEM characterization. The atomic ratio between Ag and Te was determined by energy-dispersive X-rayspectroscopy in a scanning electron microscopy. The width and thickness of the nanoribbons were determined by scanning electron microscopy and atomic force microscopy, respectively.

**Device fabrication.** Nanoribbons were sonicated into the IPA solution and then transfer to the 300 nm p-type (100) Si/SiO$_2$ substrates. Photolithography was used to pattern electrodes on the nanoribbons. Cr/Au (5 nm/120 nm) contacts were deposited in a magnetron sputtering system with a base pressure of $1 \times 10^{-8}$ torr.

**Transport measurements.** Standard DC (for large nanoribbons) and lock-in technique (for thin nanoribbons) were employed to perform four-terminal magnetoresistance measurements in a 9 Tesla Quantum Design PPMS system.

**Acknowledgements**

L. W. acknowledges the support from Singapore National Research Foundation (RCA-08/018) and MOE Tier 2 (MOE2010-T2-2-059). S. Q. S. thanks the Research Grant Council of Hong Kong under Grant No. HKU705110P. W. G. Z. acknowledges the Singapore A*STAR SERC 102 101 0019. T. Y. acknowledges NTU-SUG M4080513.

**Author contributions**

A. S., P. R. and L. W. conceived and designed the experiments. L. S., Q. Y. W. and Y. P. F. performed the band structure calculation. A. S., B. X., Q. H. L., S. Y. Z. and M. Y. H. carried out the synthesis and structure characterization. A. S., Z. P. L., C.Y. Q, T. Y., and W. G. Z. carried out the device fabrication. A. S., P. R. and B. X. performed the transport measurements. A. S., P. R., S. Y. Z., L. S., Y. P. F., S. Q. S, and L. W. contributed to the theoretical analysis and the preparation of the manuscript.

**Additional information**

**Competing financial interests:** The authors declare no competing financial interests.

**Figure captions**

**Figure 1 | Band structures and spin texture.** The calculated surface band structure of the $β$-$Ag_2Te$ film with the projected bulk band structure in the background (blue) to the 2D Brillouin zone whose shape is shown in the upper inset. The surface topological states are highlighted by the red lines. The bottom inset shows the expectation values of spin operator ($S_z$) of the surface band in $β$-$Ag_2Te$ along $k_x$ and $k_y$ directions in momentum space. Three constant-energies, which cut through the surface state below the Dirac point, are plotted. The red (blue) color means the out-of-plane components pointing out-(in-) ward of the plane. The color scale, with blue and red, indicates the intensity of negative and positive values representing the projection $S_z$.

**Figure 2 | Crystal structure of $Ag_2Te$**. (a) TEM image of an as prepared nanoribbons with the inset showing the corresponding SAED pattern. (b) Transmission electron diffraction pattern simulated based on monoclinic $Ag_2Te$ single crystal with cell parameters $a$ = 8.164 Å, $b$ = 4.468 Å, $c$ = 8.977 Å, space group $P2_1/c$ (ICSD file number: 073402). The electrons are directed to the sample at zone axis [4 $\bar{8}$ 5] with energy at 300 KeV. (c) High resolution TEM image of a focused ion beam milled $Ag_2Te$ nanoribbon with the inset showing the corresponding fast Fourier transform pattern. (d) EDS mapping for $Ag_2Te$ nanoribbon.

**Figure 3 | Four contact devices.** (a) Schematic diagram of four contact devices used in our transport experiments. (b) A scanning electron microscopy image of device in our experiments. The image shows one nanoribbon and five gold contacts on the nanoribbon.

**Figure 4 | Aharonov-Bohm oscillation of a Ag$_2$Te nanoribbon.** 1. (a) Normalized magnetoresistance of a β-Ag$_2$Te nanoribbon (sample C) with an applied magnetic field parallel to the current flowing direction at temperature 2 K, 4 K, 6 K, 8 K, and 10 K, respectively. A clear resistance oscillation with a period of 0.227 Tesla (h/e) is observed, as shown by the dotted lines. The arrows indicate a weak oscillation with a period of 0.113 Tesla (h/2e). (b) The resistivity oscillations of sample C at 2 K in a field from 0 Tesla to 9 Tesla. The arrow indicates the minimums of the oscillations with a period of h/2e. (c) The temperature dependence of the quantum oscillation amplitude. (d) Schematic diagram of Aharonov-Bohm effect measurements.

**Figure 5 | Temperature dependence of resistance.** The temperature dependence of normalized resistance at zero magnetic field of (a) sample A, (b) sample B and (c) sample C, respectively.

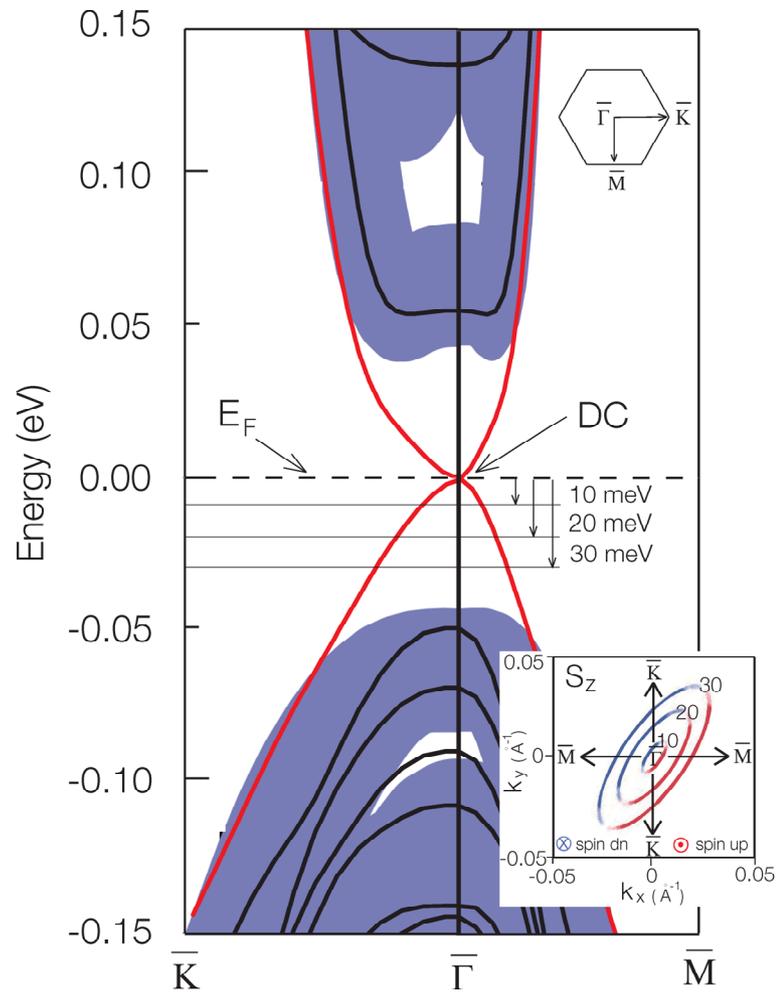

Figure 1

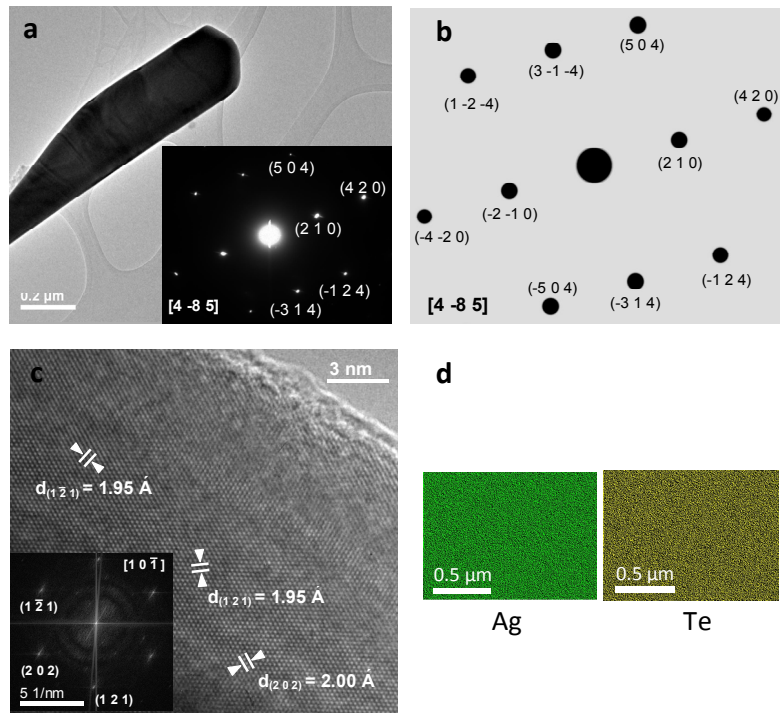

Figure 2.

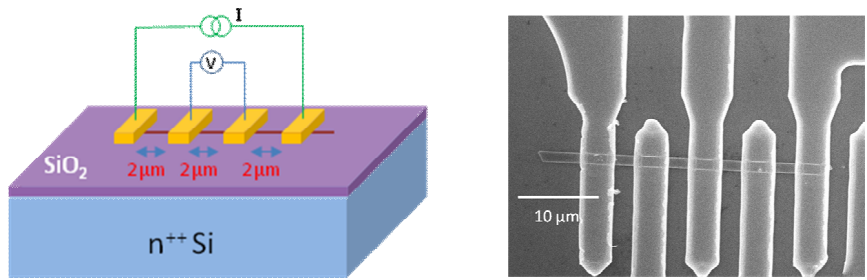

Figure 3.

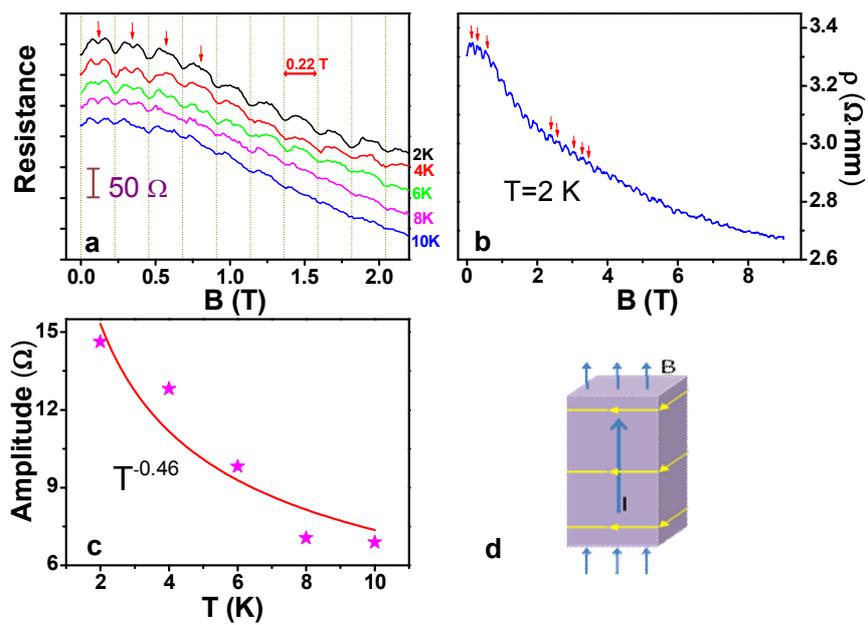

Figure 4.

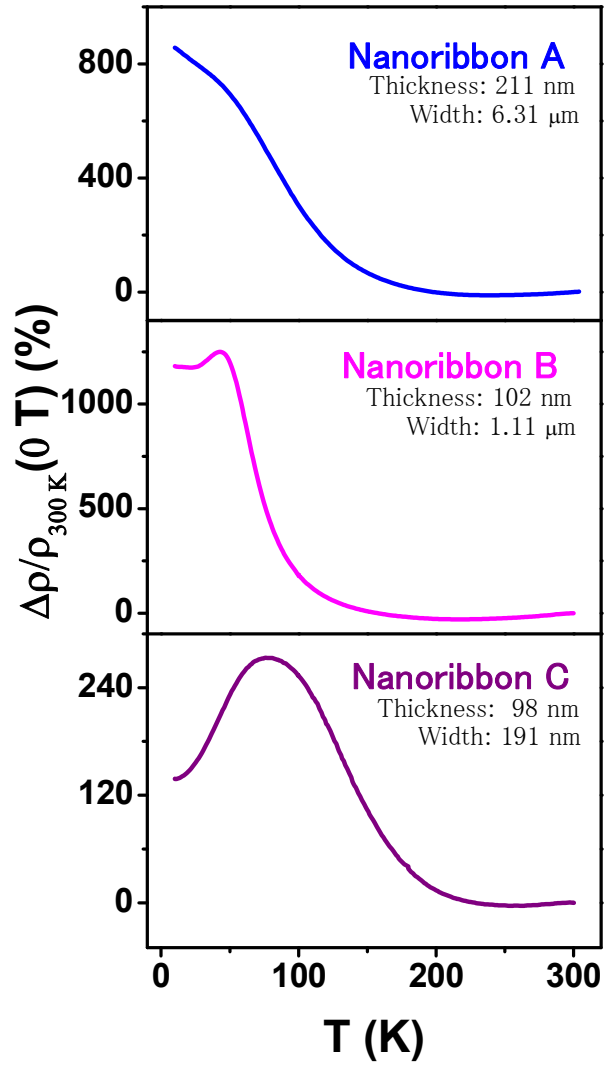

Figure 5.